\documentclass[aps,prx,showpacs,floatfix,superscriptaddress,longbibliography]{revtex4-1}
\usepackage{amsmath}
\usepackage{amssymb}
\usepackage{amstext}
\usepackage{amsopn}
\usepackage{amsfonts}
\usepackage{amsxtra}
\usepackage[english]{babel}
\usepackage{graphicx}
\usepackage{float}
\usepackage{subfigure}
\usepackage{bm}
\usepackage{multirow}
\usepackage{mathrsfs}
\usepackage{mathtools}
\usepackage{dcolumn}
\usepackage{upgreek}
\usepackage{ulem}

\usepackage[colorlinks,linkcolor=blue,urlcolor=blue,citecolor=blue]{hyperref}
\usepackage{color}
\usepackage{url}
\usepackage{breakurl}
\usepackage{relsize}

\begin{document}

\title{Robust spin order and fragile charge order in Na$_{0.5}$CoO$_2$ as revealed by time-resolved terahertz spectroscopy
}

\author{X. Y. Zhou}
\email{xinyuzxy@pku.edu.cn}
\affiliation{International Center for Quantum Materials, School of Physics, Peking University, Beijing 100871, China}

\author{S. J. Zhang}
\affiliation{Institute of Physics, Chinese Academy of Sciences, Beijing 100080, China}

\author{D. Wu}
\affiliation{Beijing Academy of Quantum Information Sciences, Beijing 100913, China}

\author{H. Wang}

\affiliation{International Center for Quantum Materials, School of Physics, Peking University, Beijing 100871, China}

\author{B. H. Li}
\author{S. F. Wu}
\affiliation{Beijing Academy of Quantum Information Sciences, Beijing 100913, China}

\author{Q. M. Liu}
\author{T. C. Hu}
\author{R. S. Li}
\author{J. Y. Yuan}
\author{S. X. Xu}
\author{Q. Wu}
\author{L. Yue}
\author{T. Dong}
\affiliation{International Center for Quantum Materials, School of Physics, Peking University, Beijing 100871, China}

\author{N. L. Wang}
\email{nlwang@pku.edu.cn}
\affiliation{International Center for Quantum Materials, School of Physics, Peking University, Beijing 100871, China}
\affiliation{Beijing Academy of Quantum Information Sciences, Beijing 100913, China}
\affiliation{Collaborative Innovation Center of Quantum Matter, Beijing, China}

%

%
\begin{abstract}
Near-infrared (NIR) pump-terahertz (THz) probe spectroscopy is used to investigate the charge and spin exciations in a strongly correlated electron compound Na$_{0.5}$CoO$_2$. This compound exhibits a coexistence of various charge and spin orders arising from intricate interactions among charge, spin, and orbital degrees of freedom. NIR pulses create significantly diverse effects on the charge and spin orders; while the charge order is easily melted, coherent magnon excitations are present in all fluences examined. Furthermore, a novel $\pi$ phase shift of the coherent magnon oscillations is observed in the pump-induced change of the terahertz electric field between regions of increasing and decreasing field change. These results unequivocally illustrate that ultrashort laser pulses enable the disentanglement of different interactions within complex systems characterized by multiple orders, providing a fresh perspective on the interplay between itinerant and localized electrons within the Co 3d t$_{2g}$ multiplets.

\end{abstract}

\maketitle
%
%
\section*{Introduction}

Ultrashort laser pulses are extensively utilized in perturbing quantum materials, facilitating the investigation of relaxation dynamics, initiation of coherent oscillations of collective modes, and creation of novel properties and phases beyond equilibrium \cite{WOS:000380759400001,RN1116,RN1333}. In the majority of cases, these pulses have been used to manipulate a specific order of material, such as the disruption of electronic order in a charge density wave (CDW) compound or the demagnetization of a magnetic compound \cite{PhysRevLett.105.187401,PhysRevLett.85.844,PhysRevLett.95.167401}. Nevertheless, in complex solids with strongly correlated electrons, intricate interactions involving spin, charge, lattice, and orbital degrees of freedom can lead to the simultaneous presence of multiple orders. Traditional spectroscopic techniques face challenges in disentangling different interactions or detecting the relative strength of coupling between different orders under equilibrium conditions. Ultrafast spectroscopy has thus become indispensable for elucidating the hierarchy or strengths of interactions in the time domain, as it allows for the monitoring of the disruption of various orders with varying excitation fluences.

\begin{figure*}
  \centering
  \includegraphics[width=8cm]{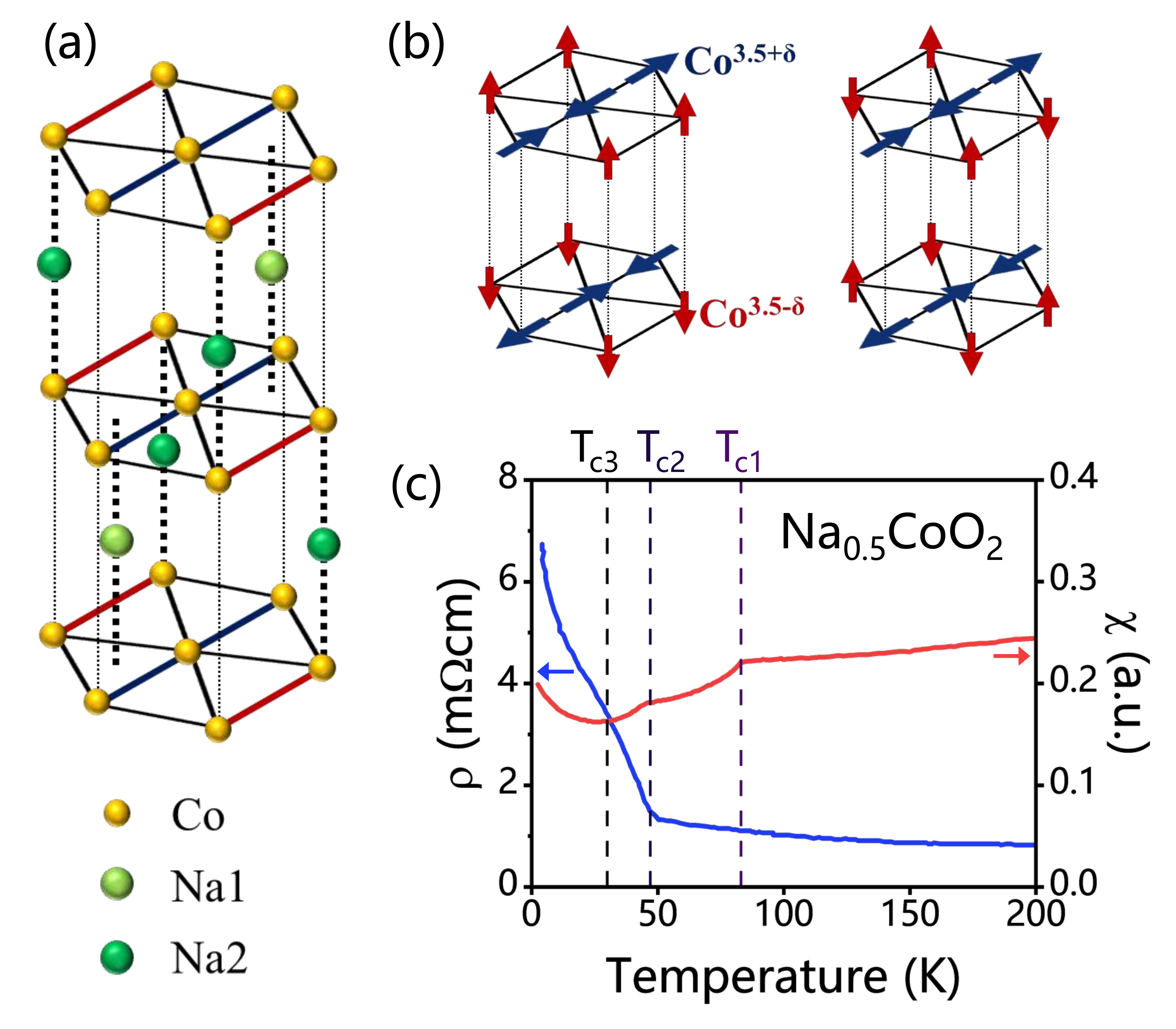}
  \caption{(a) Crystal structure of Na$_{0.5}$CoO$_2$, with oxygen sites undisplayed. The ordered occupation of Na ions leads to inequivalent Co valent states of Co$^{3.5+\delta}$ and Co$^{3.5-\delta}$ within each CoO$_2$ layer. (b) The Co$^{3.5+\delta}$ ion has larger magnetic moment, while the Co$^{3.5-\delta}$ ion has smaller magnetic moment. They form antiferromagnetic orders at T$_{c1}$ and T$_{c3}$, respectively, with two possible configuration.(c) Temperature dependent in-plane resistivity and magnetic susceptibility of Na$_{0.5}$CoO$_2$ measured by Quantum Design physical property measurement system. }
  \label{Fig:1}
\end{figure*}

Among various strongly correlated electron systems, the two-dimensional triangular lattice system Na$_x$CoO$_2$ is a particularly attractive example \cite{WOS:000181343100033,WOS:000183012000038,PhysRevB.68.020503,QHuang_2004,PhysRevLett.92.197201,PhysRevLett.92.246402,PhysRevLett.92.247001,PhysRevLett.93.147403,PhysRevB.69.180506,PhysRevLett.94.206401,PhysRevLett.100.086405,PhysRevLett.93.187203,PhysRevLett.93.237007,PhysRevLett.96.046407,PhysRevB.73.014523,PhysRevB.74.172507,PhysRevLett.101.066403}. This system comprises alternating layers of Na and CoO$_2$, with the CoO$_2$ layers structured by edge-sharing CoO$_6$ octahedra. The energy levels of the Co e$_g$ and t$_{2g}$ orbitals are differentiated by approximately 2 eV due to the oxygen octahedral crystal field, with the Fermi level falling within the Co t$_{2g}$ multiplet, which is further split by the trigonal crystal field. The properties of Na$_x$CoO$_2$, primarily governed by the Co t$_{2g}$ electrons, exhibit significant sensitivity to the Na concentration x, varying from antiferromagnet or strongly correlated Curie-Weiss metallic state in the Na-rich region (x$>$0.5) to an uncorrelated Pauli-like paramagnetic metal in the Na-poor region (x around 0.3). Superconductivity at 4.5 K is observed for x near 0.3 when water is intercalated between the CoO$_2$ layers \cite{WOS:000181343100033}. The most intriguing composition is x=0.5, where a variety of charge and spin orders are formed with a change of temperature \cite{QHuang_2004,PhysRevLett.92.247001,PhysRevLett.93.147403,PhysRevLett.100.086405,doi:10.1143/JPSJ.74.3046,PhysRevLett.96.046407,PhysRevLett.94.026403,PhysRevLett.96.046403,PhysRevB.74.172507}. At x=0.5, the Na ions are chemically ordered to form zigzag chains in an orthorhombic superstructure above room temperature, leading to inequivalent Co valent states of Co$^{3.5+\delta}$ and Co$^{3.5-\delta}$ within each CoO$_2$ layer, as shown in Fig. \ref{Fig:1} (a) and (b). The Co$^{3.5+\delta}$ ion possesses a larger magnetic moment, while the Co$^{3.5-\delta}$ ion has a smaller magnetic moment. As the temperature decreases, an antiferromagnetic (AFM) order is established at T$_{c1}\approx$ 87 K, being attributed to the ordering from Co$^{3.5+\delta}$ with magnetic moments oriented within the ab-plane. At T$_{c2}\approx$ 50 K, a charge order develops, resulting in an insulating ground state. Close to T$_{c3}\approx$ 27 K, another AFM ordering emerges from Co$^{3.5-\delta}$ with magnetic moments perpendicular to the ab-plane, causing a change in the curvature of resistivity \cite{doi:10.1143/JPSJ.74.3046,PhysRevLett.94.026403,PhysRevLett.96.046403}. The characteristics linked to spin and charge orders at these temperatures are evident in in-plane resistivity and magnetic susceptibility measurements, as depicted in Fig. \ref{Fig:1} (c). Two possible AFM orders with different configurations from the Co$^{3.5-\delta}$ ions, as displayed in Fig. \ref{Fig:1} (b), were proposed based on neutron diffraction and NMR measurements \cite{doi:10.1143/JPSJ.74.3046}. Despite being identified for two decades, the relative coupling strengths that give rise to these orders remain unclear, primarily due to the lack of experimental probes. The advent of ultrafast spectroscopy offers a promising avenue for elucidating these interaction strengths. However, to the best of our knowledge, ultrafast spectroscopy techniques have not yet been applied to this intriguing correlated electron system.

\begin{figure*}
  \centering
  \includegraphics[width=16cm]{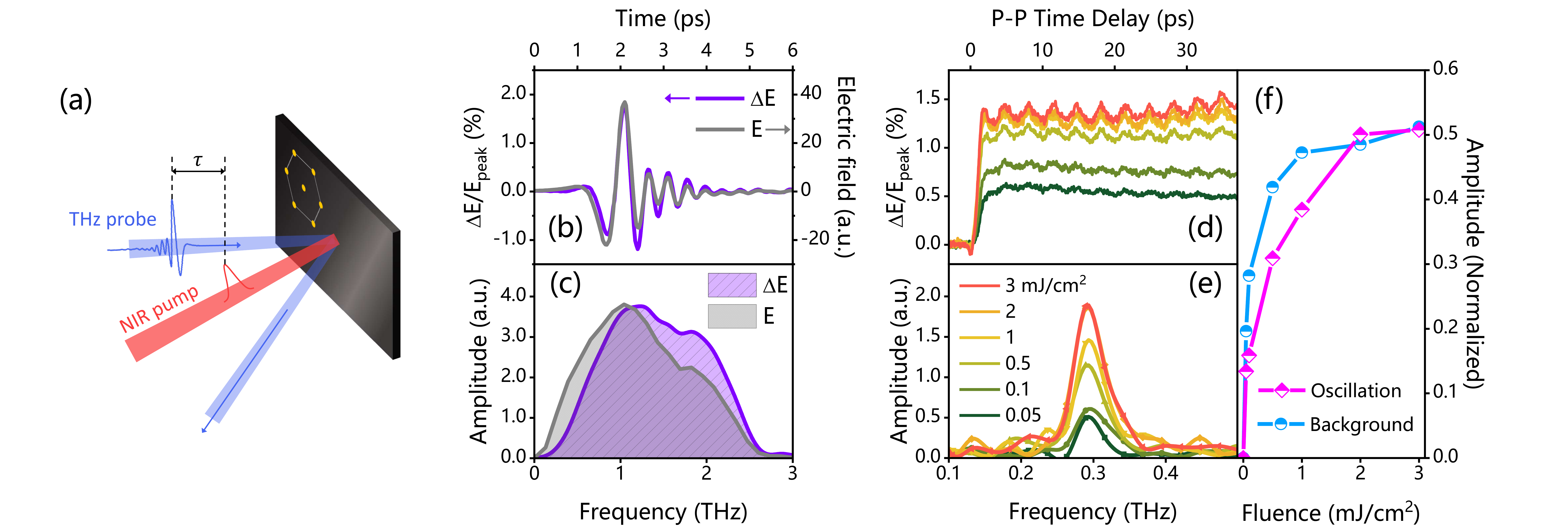}
  \caption{(a) A schematic diagram of the optical pump-terahertz probe measurement. (b) THz reflected electric field ${E}(t)$ and its variation $\Delta{E}$(t)/$E_{peak}$ measured at 10 K by double modulation and double lock-in detection. (c) The Fourier transform of (b). (d) The pump-probe signal $\Delta{E}$(t)/$E_{peak}$ at different pump fluences. (e) The Fourier transform of the oscillation extracted from the pump-probe signal at different pump fluences. (f) The fluence dependence of the oscillation strength and background of pump-probe signal. }
  \label{Fig:2}
\end{figure*}

In this work we present the near-infrared (NIR) pump terahertz (THz) probe spectroscopy study on Na$_{0.5}$CoO$_2$, with a schematic diagram in reflection geometry shown in Fig. \ref{Fig:2} (a). Upon photoexcitation by intense NIR pulses at low temperature, the conductivity in the THz region below T$_{c2}$ is greatly enhanced, yielding evidence for the ultrafast melting of charge order state. Additionally, a long-lived oscillation signal in the pump-induced change of electric field is clearly observed as a function of pump-probe time delay below magnetic T$_{c1}$. A Fourier transformation of this signal yielded a mode frequency of about 0.3 THz, which was attributed to coherent magnon mode. This marks the first precise detection of magnon excitation in the x=0.5 compound. The signal intensiy was further enhanced below T$_{c3}$ without any detectable change in frequency, and was present across all fluences applied. The results suggest that the magnetic interaction strength is much stronger than the formation of charge ordering, despite both stemming from the 3d-electrons in Co t$_{2g}$ orbitals. Furthermore, the coupling between the itinerant and localized electrons in the Co 3d t$_{2g}$ orbitals was found to be weak.

\section*{Results}

Figure \ref{Fig:2} (b) shows the static THz electric field ${E}(t)$ and its variation $\Delta{E}$(t)/$E_{peak}$ with gate time $t$ at 10 K, upon excitation by NIR pulse at 1.3 $\mu m$ with a fluence of 1 mJ/cm$^2$. Figure \ref{Fig:2} (c) displays the Fourier transformation of ${E}(t)$ and its variation $\Delta{E}$(t). It is observed that the pump-induced spectral change is more prominent at higher frequencies. Figure \ref{Fig:2} (d) displays the $\Delta{E}/E_{peak}$ slightly away from the peak position of ${E}(t)$ as a function of pump-probe (p-p) time delay $\tau$ for various excitation fluences. It is observed that a steep change occurs after excitation. A slow decaying process is observed when the fluence is below 0.5 mJ/cm$^2$, however, $\Delta{E}$/$E_{peak}$ becomes flat for fluence higher than 1 mJ/cm$^2$ within the measured pump-probe time delay $\tau$ over 40 ps. Furthermore, the signal is almost saturated beyond this fluence. These results suggest the presence of a metastable state induced by the NIR pulses. As we shall explain below based on the calculation of THz conductivity, the measurement indicates an ultrafast melting of charge order state. Very intriguingly, we observed a long-lasting coherent oscillation in the pump-probe signal $\Delta{E}$(t)/$E_{peak}$. After subtracting the step background and performing Fourier transformation, we found its central frequency to be close to 0.3 THz, as depicted in Fig. \ref{Fig:2} (e). The strength of the oscillation increased with increasing fluence, and tended to saturate when the fluence exceeded 2 mJ/cm$^2$, as illustrated in Fig. \ref{Fig:2} (f). Nevertheless, we did not observe any clear mode characteristics in the static THz spectrum, likely due to the short measurement gate time of ${E}(t)$. Since there is no optical phonon at such a low frequency, this mode is naturally attributed to the coherent magnon excitation. This assignment is further supported by the temperature-dependent measurement presented below, where the mode is only observed below the AFM transition temperature at T$_{c1}$. Notably, this research represents the first precise detection of the magnon mode in this x=0.5 composition, despite the compound being well-known to the community for two decades and intensively studied.

\begin{figure}
  \centering
  \includegraphics[width=14cm]{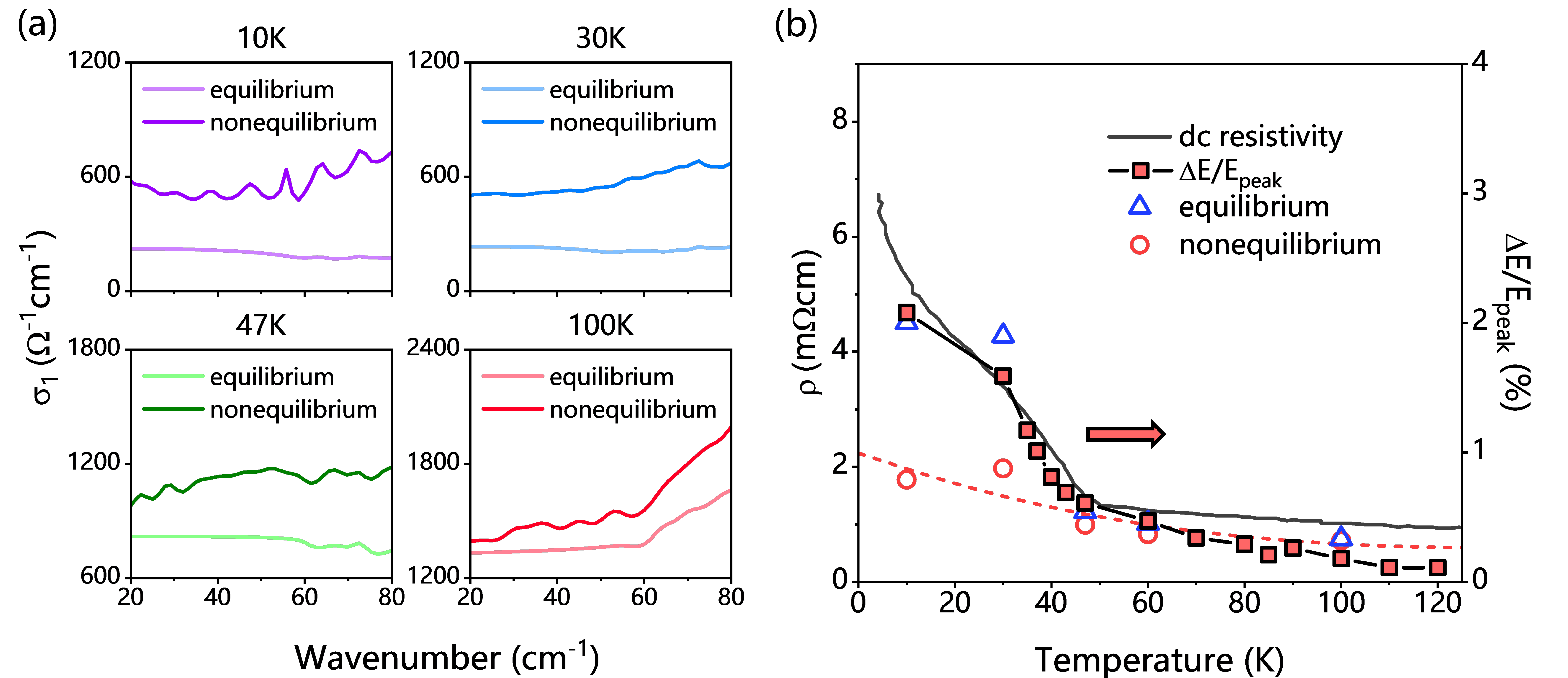}
  \caption{(a) a comparison of the static and pump-induced real part of THz conductivity $\sigma_1(\omega)$ at several selected temperatures. NIR excitation leads to an significant increase of conductivity in the THz region. (b) The dark gray solid line shows the $dc$ resistivity measured by PPMS. The light red square with black frame shows the pump-induced $\Delta{E}$(t)/$E_{peak}$ signal changes with temperature. It coincides with the trend of $dc$ resistivity. The blue triangle shows the data from equilibrium infrared conductivity spectra extracted at 20 cm$^{-1}$, The light red circle shows the pump-induced data from the THz conductivity at 20 cm$^{-1}$. The red dashed line is a guide for the eyes.}
  \label{Fig:3}
\end{figure}

In our experiment, the penetration depth of the NIR pump beam was much shorter than that of the THz probe beam, resulting in inhomogeneous excitation. To account for this, we employed a thin film model to derive the optical conductivity in the pumped region. This model assumes that the sample is uniformly excited within the penetration depth of the pump beam, while the excitation of the deeper part of the sample is neglected. The calculation method and rationale of the model have been presented elsewhere \cite{PhysRevX.10.011056,PhysRevB.98.020506}. Figure \ref{Fig:3} (a) shows a comparison of the static and pump-induced real part of THz conductivity $\sigma_1(\omega)$ by 1 mJ/cm$^2$ at several different temperatures. The equilibrium conductivity was obtained from the Kramers-Kronig transformation of the infrared reflectance measured by Fourier transform infrared spectroscopy. We observed a significant increase of conductivity in the THz region after photoexcitation, particularly at low temperatures. We compared the $dc$ resistivity and the pump-induced THz conductivity, as well as the measured $\Delta{E}$(t)/$E_{peak}$. As shown in Fig. \ref{Fig:3} (b), the compound underwent a charge order transition at T$_{c2}$, where the resistivity increased sharply below T$_{c2}$. Similarly, $\Delta{E}$(t)/$E_{peak}$ followed the same trend. This resulted in a sharp increase of THz conductivity after excitation. The pump-induced resistivity, e.g. at 20 cm$^{-1}$, also dropped significantly below T$_{c2}$, following the temperature-dependent resistivity trend established above T$_{c2}$. The results undoubtedly suggest a melting of charge order by ultrashort NIR laser pulses. Nevertheless, the conductivity spectrum does not appear to be exactly Drude-like in the measured limited THz frequency range. This is unsurprising, as even above T$_{c2}$, the $dc$ resistivity still decreases slightly when temperature increases.

\begin{figure}
  \centering
  \includegraphics[width=14cm]{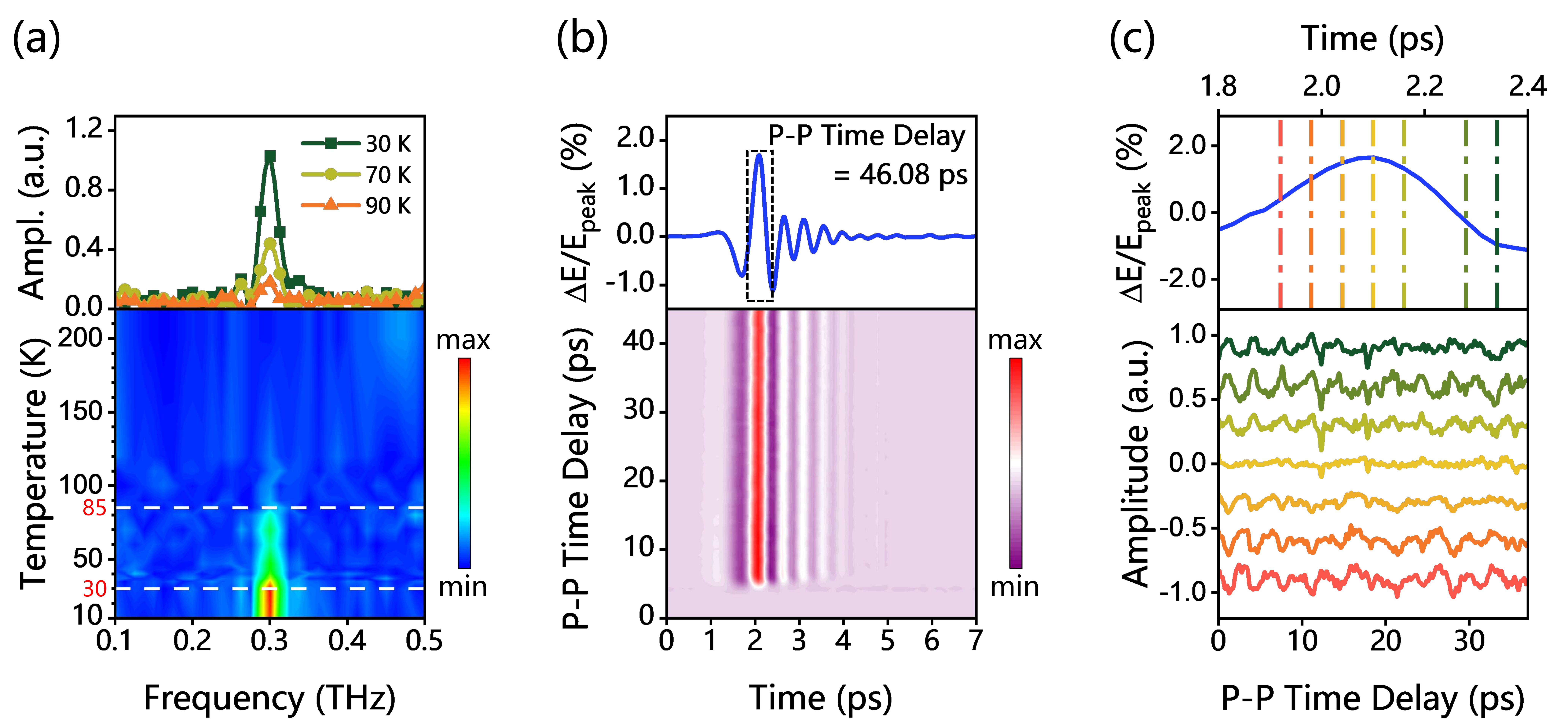}
  \caption{ (a) Upper panel: the coherent magnon mode at several temperatures. Lower panel: the intensity plot of the magnon mode as a function of temperature. The mode appears below T$_{c1}$ and its intensity is further enhanced below T$_{c3}$. (b) Upper panel: $\Delta{E}$(t)/$E_{peak}$ intercepts along pump-probe time delay $\tau$= 46.08 ps. Lower panel: Two-dimensional scanning of $\Delta{E}(t,\tau)$/$E_{peak}$ signal along $t$ and $\tau$. (c) Upper panel: an enlarged region along gate time near main peak (marked by black dotted box) displayed in the upper panel of (b). Lower panel: oscillations after subtracting background at different gate times labelled in the upper panel. The curves were shifted vertically for clarity. There is a $\pi$-phase shift of the oscillations between the region when $\Delta{E}$/$E_{peak}$ is increasing or decreasing. At the gate time when $\Delta{E}$ reaches the maximum, the oscillation becomes invisible.}
  \label{Fig:4}
\end{figure}

Let us now focus on the long-lasting coherent oscillation in $\Delta{E}$(t)/$E_{peak}$ observed along the pump-probe time delay $\tau$ (Fig. \ref{Fig:2} (d)). Its central frequency is approximately 0.3 THz. The upper panel of Fig. \ref{Fig:4} (a) displays the mode at several representative temperatures, while the lower panel of Fig. \ref{Fig:4} (a) shows the intensity plot of detailed temperature-dependent evolution of this mode. It is evident that the mode is formed below the first magnetic phase transition temperature T$_{c1}$, which is associated with the antiferromagnetic order of Co$^{3.5+\delta}$, and represents a coherent magnon excitation. The intensity does not show any significant change until the second magnetic phase transition near T$_{c3}$, which is due to the antiferromagnetic order of Co$^{3.5-\delta}$. Notably, the mode frequency shows no softening as T$_{c1}$ is approached, likely due to the dominating energy scale of the antiferromagnetic interaction over temperature changes. In addition, the second magnetic phase transition, associated with the ordering of the smaller moments of Co$^{3.5+\delta}$, only alters the mode strength, not the frequency. It worth mentioning that the observed magnon frequency is in line with the neutron scattering measurement on Na$_x$CoO$_2$ with x=0.75 - 0.85 where AFM order forms near 22 K, in which a spin gap of 1$\sim$2 meV was detected near Brillouin zone center in ordered state \cite{PhysRevLett.94.157206,PhysRevB.73.054405}.

A more intriguing observation is that there is a $\pi$-phase shift of the coherent magnon oscillations between regions when the $\Delta{E}$(t)/$E_{peak}$ is increasing or decreasing. In the lower panel of Fig. \ref{Fig:4} (b) we plot the two-dimensional scanning of $\Delta{E}(t,\tau)$/$E_{peak}$ signal along $t$ and $\tau$. The upper panel of Fig. \ref{Fig:4} (b) shows the  $\Delta{E}$(t)/$E_{peak}$ at a specific pump-probe time delay $\tau$= 46.08 ps. The upper panel of Fig. \ref{Fig:4} (c) is an enlarged region along gate time near the main peak (marked by black dotted box) displayed in the upper panel of Fig. \ref{Fig:4} (b). Oscillations along the pump-probe delay time $\tau$ after subtracting background at different gate times labelled in the upper panel are displayed in the lower panel of Fig. \ref{Fig:4} (c). It is evident that the oscillations in the region when the $\Delta{E}$(t)/$E_{peak}$ is increasing have a phase opposite to that when the $\Delta{E}$(t)/$E_{peak}$ is decreasing. At the gate time when $\Delta{E}$(t)/$E_{peak}$ reaches the maximum, the oscillation disappears. Similar situation were observed for the gate times near other peaks in $\Delta{E}$(t)/$E_{peak}$. As far as we know, there has been no prior observation of a $\pi$-phase shift in the pump-induced change of the THz electric field, although coherent magnon has been detected in such measurements in certain magnetic compounds \cite{Belvin2021}. This observation seems to suggest that the procession of the magnetic moment seen in the pump-probe signal is driven by the magnetic component of the pump-induced change of THz wave, which is perpendicular to the change of electric component of THz wave, $\Delta{E}$(t)/$E_{peak}$. Hence, they have opposite directions when $\Delta{E}$(t)/$E_{peak}$ is increasing or decreasing, leading to opposite oscillation phases due to torque effect.

\section*{Discussion}

The above experimental results reveal unambiguously that the NIR photoexcitation in Na$_{0.5}$CoO$_2$ has significantly different effect on the charge and spin order responses. The charge order, which results in a metal-insulator like transition at T$_{c2}$, is easily melted upon NIR photoexcitation. The system was driven to a metastable metallic-like state with a saturation of pump fluence roughly beyond 1 mJ/cm$^2$. However, the coherent magnon excitation is always present over all pump fluences we used. In fact, the magnon mode strength continues to increases with increasing pump fluence and tend to saturate above 2 mJ/cm$^2$. The observation suggests that the magnetic order is always present in our measurement.

The results of this study provide novel insights into the complex interactions within Na$_x$CoO$_2$, a strongly correlated electron system with active charge, spin and orbital degrees of freedom. As is introduced in the beginning, the low-energy physical properties in Na$_x$CoO$_2$ are determined by the t$_{2g}$ manifold, which is further split into one a$_{1g}$ and two e$'$$_g$ orbitals under trigonal crystal field \cite{PhysRevB.71.205103,PhysRevLett.94.206401,PhysRevLett.96.046403,PhysRevB.69.214516}. For the composition of x=0.5, Co forms inequivalent valent states of Co$^{3.5+\delta}$ and Co$^{3.5-\delta}$ due to the ordered occupation of the Na ions, the three t$_{2g}$ multiplets are not fully occupied. Apparently, due to the competition among the kinetic energy, Coulomb interaction and Hund's coupling of electrons within the t$_{2g}$ manifold, some of the orbitals are more itinerant and contribute more to charge conduction, while others are more localized and contribute to the formation of local magnetic moments. The situation is analogous to the multi-orbital physics of Fe-based superconductors \cite{Haule_2009,Georges2013,PhysRevLett.112.177001}. The local magnetic moments from inequivalent Co sites form antiferromagnetic orders at T$_{c1}$ and T$_{c3}$, respectively, while the relatively itinerant electrons contribute to the conductivity and become localized at T$_{c2}$ when a charge order forms. The current time-resolved THz spectroscopy reveals distinct photoexcitation effect on the charge and spin orders, indicating a robust magnetic interaction between localized moments and fragile charge order susceptibility to external perturbations. Furthermore, the study also suggests a weak coupling between the itinerant and the more localized electrons in the t$_{2g}$ multiplets, as negligible mutual effect is observed.

To summarize, we employed NIR pump-THz probe spectroscopy to investigate Na$_{0.5}$CoO$_2$, a material known for its multiple charge and spin orders at low temperatures. Our research represents the first application of ultrafast spectroscopy techniques to this intriguing correlated electron system, which have provided novel insights into the spin and charge orders within the compound. Our findings reveal that intense NIR photoexcitation has a distinct impact on the charge and spin orders: the low temperature charge order is easily melted, while coherent magnon excitation, observed as a persistent oscillation in the pump-induced THz electric field change, remains present across all fluences. Moreover, the magnon does not exhibit any noticeable frequency softening near the antiferromagnetic transition temperature T$_{c1}$, and its intensity is notably enhanced below T$_{c3}$. Additionally, we observed a novel $\pi$-phase shift in the coherent magnon oscillations in regions where the rate of change of $\Delta{E}$(t)/$E_{peak}$ is either increasing or decreasing. These results also suggest a relatively weak interaction between itinerant and localized electrons in the Co 3d t$_{2g}$ manifold, offering a fresh perspective on the interplay between these electron states within the multiplets.

\section*{Materials and Methods}

\textbf{Sample preparation and equilibrium information.} Single crystals of Na$_{0.5}$CoO$_2$ were grown using the floating-zone technique \cite{PhysRevB.73.014523}. Temperature-dependent in-plane resistivity and magnetic susceptibility shown in Fig. \ref{Fig:1} (c) were measured in a Quantum Design physical property measurement system. A standard four-leads technique was used for resistivity measurement. A magnetic field of 100 Os was applied parallel to the ab-plane in the magnetic susceptibility measurement. The equilibrium infrared reflectivity spectra were acquired using a Fourier-transform infrared spectrometer (Bruker 60v) with \emph{in-situ} gold and aluminum deposition technique \cite{PhysRevLett.93.147403}. The real part of optical conductivity in equilibrium state was obtained by Kramers-Kronig transformation of reflectance spectrum.

\textbf{NIR pump-THz probe spectroscopy.} NIR pump-THz probe in reflection geometry was employed to monitor the temporal evolution of charge and spin order in response to near-infrared photoexcitation. Double modulation and double lock-in detection were employed to acquire equilibrium THz electric field ${E}(t)$ and its variation $\Delta{E}$(t)/$E_{peak}$ simultaneously. Further details of the experimental setup and measurement techniques can be found in the supplementary materials.

\section*{Acknowledgments}
This work was supported by National Natural Science Foundation of China (No.~11888101), the National Key Research and Development Program of China (No.~2022YFA1403901).

\section*{Competing interests} The authors declare that they have no competing interests.\\

\section*{Data and materials availability} All data needed to evaluate the conclusions in the paper are present in the paper and/or the Supplementary Materials.

\section*{Supplementary materials}
\section{Experimental system}

The amplified Ti:Sapphire laser provides femtosecond pulses centered at 800 nm with 35 fs pulse duration and 1 kHz repetition rate, the majority power injects into OPA to generate NIR-pump laser with 1300 nm centerwavelength. And the rest pulses are divided into 2 parts, one part is incidents into a ZnTe crystal to generate THz wave, the other part called sampling light is confocal with the reflected THz-probe electric field onto another ZnTe crystal to detect THz or pump-induced change signal. NIR-pump light and THz-probe are modulated by a chopper with frequency of 250 and 500 Hz respectively and are synchronized with the laser. All the THz light paths are settled in a sealed chamber and injected with nitrogen continuously to reduce the absorption by water vapor.

\begin{figure*}[h]
	\centering
	\includegraphics[width=12cm]{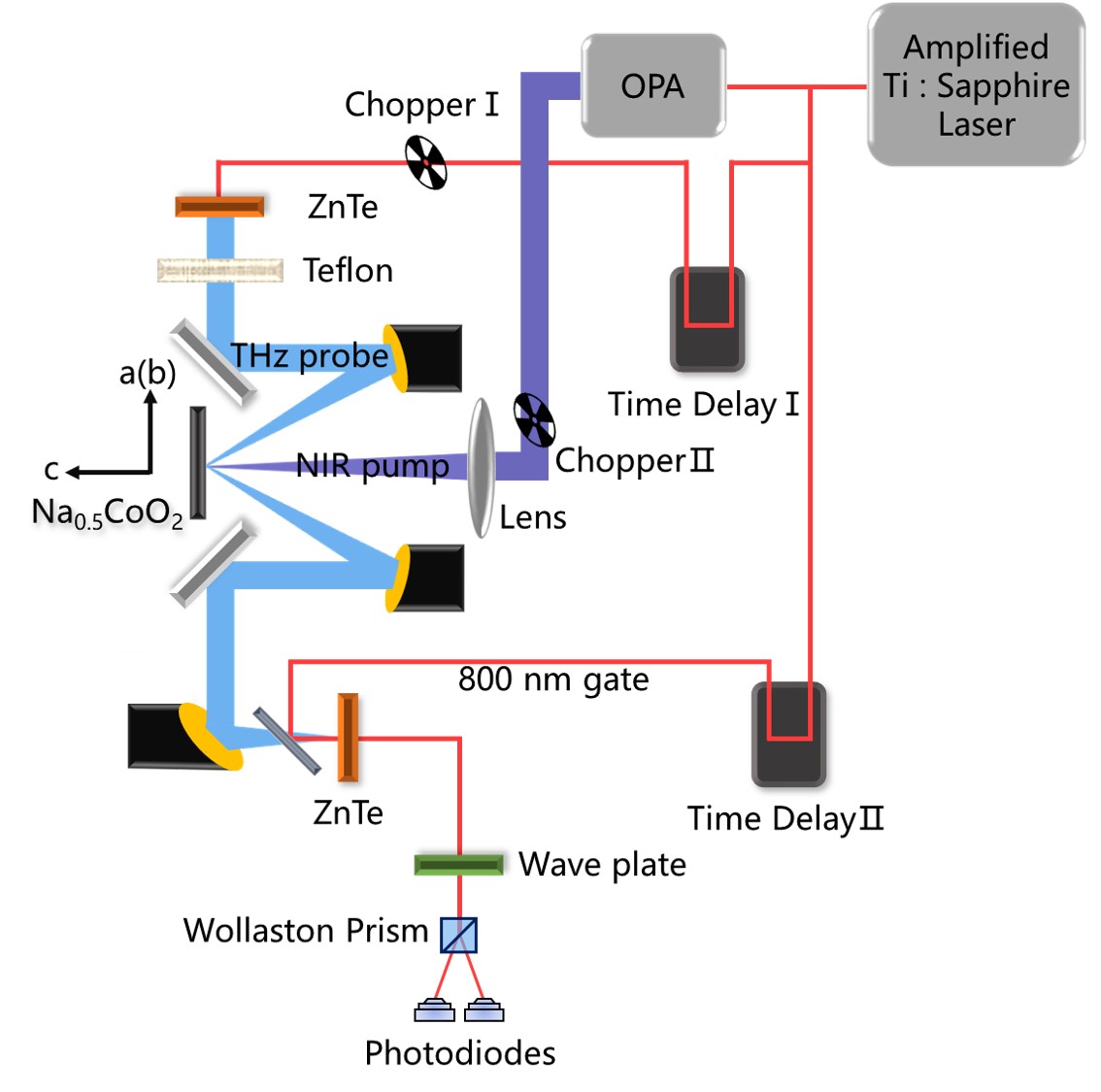}\\
	\caption{The schematic diagram of NIR-pump THz-probe reflection spectral system.}\label{Fig:5}
\end{figure*}

There are two mechanical delay devices in THz generation and sampling light path respectively (shown in Fig. \ref{Fig:5}). Therefore the NIR-pump THz-probe experimental system has two time dimensions, THz sampling time t (Time Delay I) and pump-probe delay time $\tau$ (Time Delay II). In this system, THz probe waves strike the specimen  at 30$^\circ$. To ensure the detecting region is excited uniformly, the diameter of THz-probe beam is 600-700 $\mu$m, and the diameter of NIR-pump beam is more than 1 mm. The THz-probe wave is generally set perpendicular to the plane of propagation, that is, the transverse electrical mode (TE wave).

\section{Detection technique}

\begin{figure*}[h]
	\centering
	\includegraphics[width=16cm]{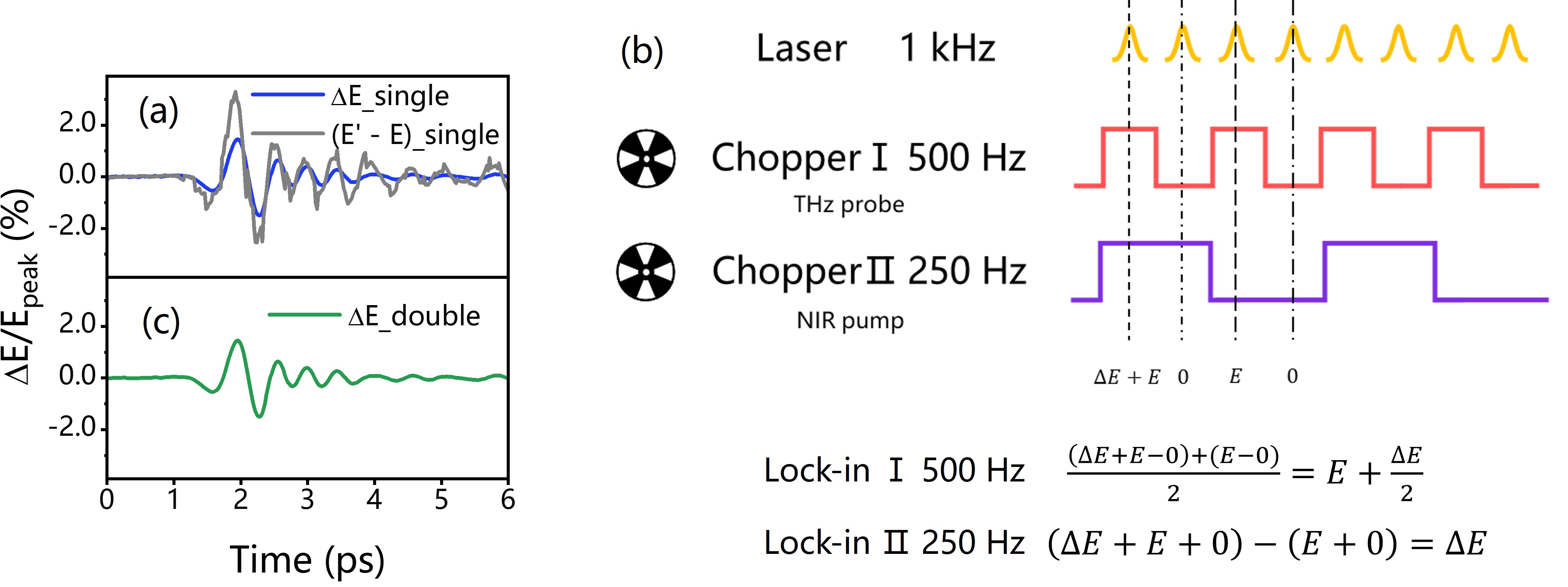}\\
	\caption{(a) The pump-induced THz-probe electric field measured by single modulation and lock-in detection. (b) Schematic diagram of double modulation and lock-in detection. (c) The pump-induced THz-probe electric field measured by double modulation and lock-in detection}\label{Fig:6}
\end{figure*}

We need to determine the phase of the nonequilibrium signal by single modulation and single lock-in detection as shown in Fig. \ref{Fig:6}(a), E'-E depicts the difference value between NIR-pump induced and equilibrium reflected THz-probe electric field, and this value should be the same of the pump induced differential electric field $\Delta$E. Once the phase is determined, E and $\Delta$E can be obtained simultaneously by means of double modulation and double lock-in detection shown in Fig. \ref{Fig:6}(b). $\Delta$E measured by double modulation and single modulation , shown in Fig. \ref{Fig:6}(c) and (a) respectively, are identical in both amplitude and phase.

\section{Penetration depth}

\begin{figure*}[h]
	\centering
	\includegraphics[width=8cm]{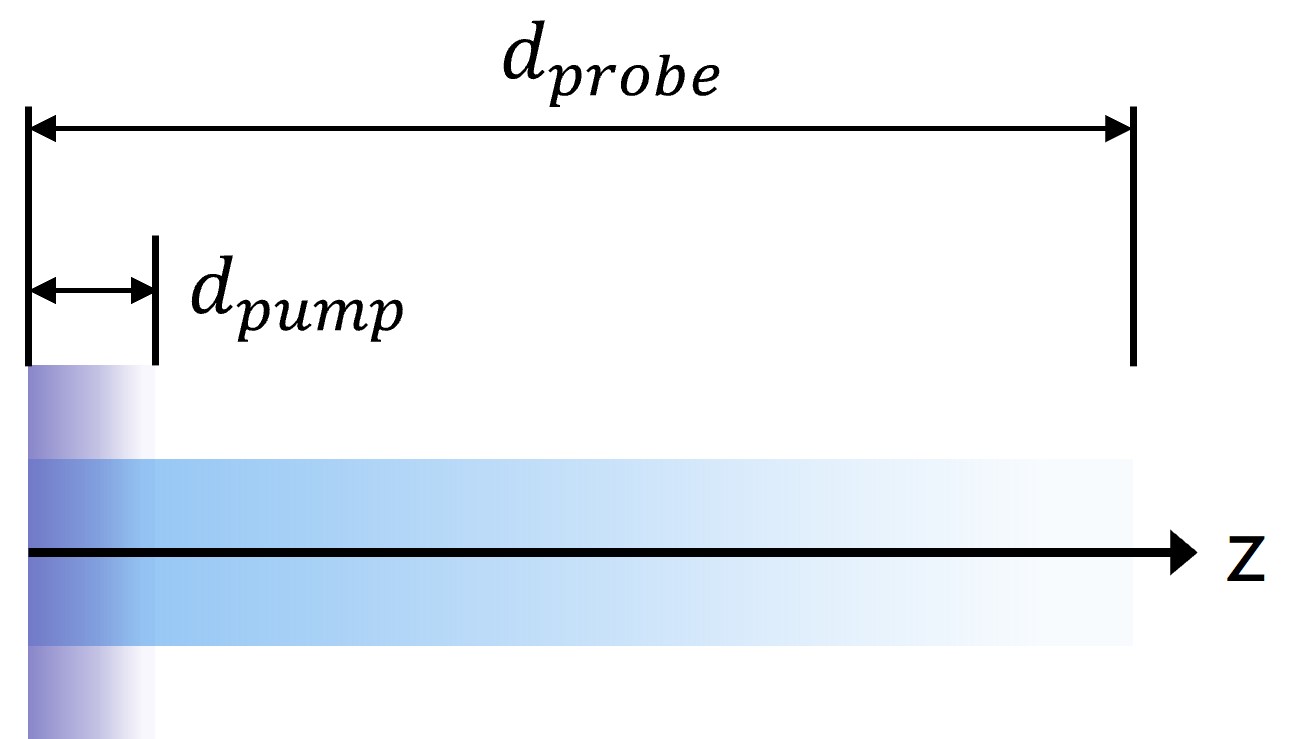}\\
	\caption{Mismatch of NIR-pump and THz-probe light.}\label{Fig:7}
\end{figure*}

In our experimental configuration, the penetration depth of NIR-pump light d$_{pump}$ = 0.31 $\mu$m is far less than THz-probe light d$_{probe}$ = 4.5 $\mu$m, this corresponds to a scenario of inhomogeneous excitation (shown in Fig. \ref{Fig:7}), only a vey small part of probe-covered regime contributes the pump-induced signal. To describe the penetration depth mismatch of pump and probe light, we simplify the situation to a thin film model \cite{PhysRevX.10.011056}. In this model, the specimen is assumed to be excited uniformly within the penetration depth of the pump light while the excitation of the deeper part can be neglected. All the depths are calculated by equilibrium data from Fourier Transform infrared spectroscopy.

\section*{References}

\bibliography{NCO}

\end{document}